\documentclass[bib]{ba}

\pdfoutput=1

\usepackage{bayes}

\usepackage{amsfonts,amssymb,amsmath,amscd,amsthm,latexsym}
\usepackage{graphicx}

\newtheorem{example}{Example}
\newenvironment{ex}
  {\begin{example}\begin{em}}
  {\end{em}\hfill$\blacktriangleleft$\end{example}}

\newcommand{\dd}{\,\text{d}}

\begin{document}

\inserttype[ba0001]{article}
\author{Robert, C.P., and Marin, J.-M.}{%
  Christian P.~Robert\footnote{xian@ceremade.dauphine.fr}\\CEREMADE, Universit\'e Paris Dauphine, and CREST, INSEE
  \and
  Jean-Michel Marin\footnote{jean-michel.marin@inria.fr}\\INRIA Saclay, Universit\'e Paris Sud, Orsay, and CREST, INSEE
}
\title[Difficulties with an approximation technique]{On some difficulties with a posterior probability approximation technique}

\maketitle

\begin{abstract}
In \cite{scott:2002} and \cite{congdon:2006}, a new method is advanced to compute posterior probabilities of models
under consideration. It is based solely on MCMC outputs restricted to single models, i.e., it is bypassing reversible
jump and other model exploration techniques. While it is indeed possible to approximate posterior probabilities
based solely on MCMC outputs from single models, as demonstrated by \cite{gelfand:dey:1994} and
\cite{bartscami}, we show that the proposals of \cite{scott:2002} and \cite{congdon:2006} 
are biased and advance several arguments towards this thesis, the primary one being the confusion between
model-based posteriors and joint pseudo-posteriors. 
From a practical point of view, the bias in Scott's (2002) approximation appears to be much more severe than the one in
Congdon's (2006), the latter being often of the same magnitude as the posterior probability it approximates, although
we also exhibit an example where the divergence from the true posterior probability is extreme.

\end{abstract}

\section{Introduction}
Model selection is a fundamental statistical issue and a clear asset of the Bayesian methodology
but it faces severe computational difficulties because of the requirement to explore simultaneously the parameter spaces
of all models under comparison accurately enough to provide sufficient approximations to the posterior probabilities
of all models. When \cite{green:1995} introduced reversible jump techniques, it was perceived by the
community as the second MCMC revolution in that it allowed for a valid and efficient exploration of the
collection of models and the subsequent literature on the topic exploiting reversible jump MCMC
is a testimony to the appeal of this method. Nonetheless, the implementation of reversible jump techniques 
in complex situations may face difficulties
or at least inefficiencies of its own and, despite some recent advances in the devising of the jumps underlying
reversible jump MCMC \citep{brooks:giudici:roberts:2003}, the care required in the construction of those
jumps often acts as a deterrent from its applications. 

There are practical alternatives to reversible jump MCMC when the number of models under consideration is small
enough to allow for a complete exploration of those models. Integral approximations using importance
sampling techniques like those found in \cite{gelfand:dey:1994}, based on a harmonic mean representation
of the marginal densities, and in \cite{gelman:meng:1998}, focussing on the optimised selection of the
importance function, are advocated as potential solutions, see \cite{chen:shao:ibrahim:2000} for a detailed introduction.
The reassessment of those methods by \cite{bartscami} showed the connection between a virtual reversible jump MCMC
and importance sampling \citep[see also][]{chopin:robert:2007}. In particular, those papers 
demonstrated that the output of MCMC samplers on each single model could be used to produce approximations 
of posterior probabilities of those models, via some importance sampling methodologies also related
to \cite{newton:raftery:1994}.

In \cite{scott:2002} and \cite{congdon:2006}, a new and straightforward method is advanced to 
compute posterior probabilities of models under scrutiny based solely on MCMC outputs restricted 
to single models.  While this simplicity is quite appealing for the approximation of those probabilities, we believe 
that both proposals of \cite{scott:2002} and \cite{congdon:2006} are inherently biased and 
we advance in this note several arguments towards this 
thesis. In addition, we notice that, to overcome the bias we thus exhibited,
a valid solution would call for the joint simulation of parameters under all 
models (using priors or pseudo-priors)
and, in this step, the primary appeal of the methods would thus be lost compared to
the one proposed by \cite{carlin:chib:1995}, from which both \cite{scott:2002} 
and \cite{congdon:2006} are inspired.

We want to point out at this stage that the original purpose of \cite{scott:2002} is to provide a survey of Bayesian
methods for the analysis of hidden Markov models and thus that the approximation we analyse here is introduced as a side
remark within the whole paper. If we insist here on the bias produced by Scott's (2002) approximation, it is because
it generated followers, including \cite{congdon:2006}, and because both approximations are based on the same erroneous
interpretation of the marginal distribution in Bayesian model choice. We also note that Congdon's (2006) approximation
often produces values that are numerically of the same magnitude as the true value of the posterior probabilities, 
with sometimes very close proximity as illustrated in Example 2 of Section \ref{sec:ex}, but also potential 
severe mishaps as in Example \ref{ex:Ex} of Section \ref{sec:ex}.

\section{The methods}
In a Bayesian framework of model comparison \citep[see, e.g.,][]{robert:2001},
given $D$ models in competition, $\mathfrak{M}_k$, with densities $f_k(y|\theta_k)$, 
and prior probabilities $\varrho_k = P(M=k)$ $(k=1,\ldots,D)$, the posterior probabilities
of the models $\mathfrak{M}_k$ conditional on the data $y$ are given by
$$
P(M=k|y) \propto \varrho_k \,\int f_k(y|\theta_k) \pi_k(\theta_k)\dd\theta_k\,,
$$
the proportionality term being given by the sum of the above and $M$ denoting the unknown
model index.

In the specific setup of hidden Markov models, the solution of Scott (2002, Section 4.1) is to generate,
simultaneously and independently, $D$ MCMC chains 
$$
(\theta_k^{(t)})_t\,, \qquad 1\le k\le D\,,
$$
with stationary distributions $\pi_k(\theta_k|y)$ and to approximate $P(M=k|y)$ by
$$
\tilde{\varrho}_k(y) \propto \varrho_k \sum_{t=1}^T \left\{ f_k(y|\theta_k^{(t)}) \bigg/ 
\sum_{j=1}^D \varrho_j\,f_j(y|\theta_j^{(t)}) \right\}\,,
$$
as reported in formula (21) of \cite{scott:2002}, with the indication that 
{\em (21) averages the $D$ likelihoods corresponding to each $\theta_j$ over the life
of the Gibbs sampler (p.347),} the latter being understood as {\em independently 
sampled $D$ parallel Gibbs samplers (p.347).}

Adopting a more general perspective, the proposal of \cite{congdon:2006} for an 
approximation of the $P(M=k|y)$'s follows both from Scott's (2002) approximation
and from the pseudo-prior construction of \cite{carlin:chib:1995}
that predated reversible jump MCMC by saturating the parameter space with an artificial
simulation of all parameters at each iteration. However, due to a very special (and, we
believe, mistaken) choice of pseudo-priors discussed below, Congdon's (2006, p.349) 
approximation of $P(M=k|y)$ eventually reduces to the estimator
$$
\hat{\varrho}_k(y) \propto \varrho_k \sum_{t=1}^T \left\{ f_k(y|\theta_k^{(t)}) \pi_k(\theta^{(t)}_k)
\bigg/ \sum_{j=1}^D \varrho_j\, f_j(y|\theta_j^{(t)}) \pi_j(\theta^{(t)}_j) \right\} \,,
$$
where the $\theta_k^{(t)}$'s are samples from $\pi_k(\theta_k|y)$ (or approximate samples
obtained by an MCMC algorithm). 
This is a simple and readily implementable formula that attracted other researchers like \cite{chen:gerlach:so:2008}.

Although both approximations $\tilde{\varrho}_k(y)$ and $\hat{\varrho}_k(y)$ 
clearly differ in their expressions, by the addition of a $\pi_k(\theta^{(t)}_k)$ term
in Congdon's (2006) formula, they fundamentally relate to the same notion that parameters from other
models can be ignored when conditioning on the model index $M$. This approach is therefore bypassing
the simultaneous exploration of several parameter spaces and it restricts the simulation effort to
marginal samplers on each separate model. This feature is very appealing since it cuts most
of the complexity from the schemes both of \cite{carlin:chib:1995} and of \cite{green:1995}.
We however question the foundations of those approximations as presented in 
both \cite{scott:2002} and \cite{congdon:2006} and advance below arguments that both authors
are using incompatible versions of joint distributions on the collection of parameters that
jeopardise the validity of the approximations.

\section{Difficulties}

The sections below expose the difficulties found with both methods, following the 
arguments advanced in \cite{scott:2002} and \cite{congdon:2006}, respectively. The fundamental difficulty 
with both approaches appears to us to stem from a confusion between the model dependent 
simulations and the joint simulations based on a pseudo-prior scheme as in \cite{carlin:chib:1995}. 
Once this difficulty is resolved, it appears that the corresponding approximation of $P(M=k|y)$ by $\hat{P}(M=k|y)$ 
does require a joint simulation of all parameters and thus that the solutions proposed in \cite{scott:2002} and
\cite{congdon:2006} are of the same complexity as the proposal of \cite{carlin:chib:1995}. 
If single models MCMC chains are to be used, alternative approaches described for instance in 
\cite{chen:shao:ibrahim:2000} and compared in \cite{gamerman:lopes:2006} can be implemented.

\subsection{Incorrect marginals}
We denote by $\theta=\left(\theta_1,\ldots,\theta_D\right)$ the collection of parameters for all models under consideration.
Both \cite{scott:2002} and \cite{congdon:2006} start from the representation
$$
P(M=k|y) = \int P(M=k|y,\theta) \pi(\theta|y) \dd \theta
$$
to justify the approximation
$$
 \hat{P}(M=k|y) = \sum_{t=1}^T P(M=k|y,\theta^{(t)}) / T \,.
$$
This is indeed an unbiased estimator of $P(M=k|y)$ provided the $\theta^{(t)}$'s are 
generated from the correct (marginal) posterior
\begin{align}
\pi(\theta|y) & = \sum_{k=1}^D P(\theta,M=k|y) \label{eq:marg} \\
              & \propto \sum_{k=1}^D \varrho_k\,f_k(y|\theta_k) \prod_j \pi_j(\theta_j) \nonumber \\
	          & = \sum_{k=1}^D \varrho_k\,m_k(y)\,\pi_k(\theta_k|y) \prod_{j\ne k} \pi_j(\theta_j)\,.
\label{eq:noway} 
\end{align}
In both papers, the $\theta^{(t)}$'s are instead simulated as independent outputs from the componentwise
posteriors $\pi_k(\theta_k|y)$ and this divergence jeopardises the theoretical validity of the approximation.
The error in both interpretations stems from the fact that, while the $\theta^{(t)}_k$'s are (correctly)
independent given the model index $M$, this independence does not hold once $M$ is integrated out, 
which is the case for the $\theta^{(t)}_k$'s in the above approximation $\hat{P}(M=k|y)$.

\subsection{MCMC versus marginal MCMC}\label{sec:marge}
When \cite{congdon:2006} defines a Markov chain $(\theta^{(t)})$ at the top of page 349, 
he indicates that the components of $\theta^{(t)}$ are made of independent Markov chains $(\theta_k^{(t)})$ 
simulated with MCMC samplers related to the respective marginal posteriors $\pi_k(\theta_k|y)$, following
the approach of \cite{scott:2002}. 
The aggregated chain $(\theta^{(t)})$ is thus stationary against the product of those marginals,
$$
\prod_{k=1}^D \pi_k(\theta_k|y)\,.
$$

However, in the derivation of \cite{carlin:chib:1995}, the model is defined in terms of \eqref{eq:marg}
and the Markov chain should thus be constructed against \eqref{eq:marg}, not against the product of the model
marginals. Obviously, in the case of \cite{congdon:2006},
the fact that the pseudo-joint distribution does not exist because of the flat prior
assumption (see Section \ref{sec:impro} for a proof) prevents this construction but, in the case
the flat prior is replaced with a proper (pseudo-) prior, 
the same statement holds: the probabilistic derivation of $P(M=k|y)$ relies on the pseudo-prior 
construction and, to be valid, it does require the completion step at the core of 
\cite{carlin:chib:1995}, where parameters {\em need} to be simulated from the pseudo-priors.
Generating from the component-wise posteriors $\pi_k(\theta_k|y)$ produces a bias.

Similarly, in \cite{scott:2002}, the target of the Markov chain $(\theta^{(t)},M^{(t)})$ should be
the distribution
$$
P(\theta,M=k|y) \propto \pi_k(\theta_k)\,\varrho_k\,f_k(y|\theta_k)\,\prod_{j\ne k} \pi_j(\theta_j)
$$
and the $\theta_j^{(t)}$'s should thus be generated from the prior $\pi_j(\theta_j)$ when 
$M^{(t)}\ne j$---or equivalently from the corresponding marginal if one does not condition on $M^{(t)}$, 
but simulating a Markov chain with stationary distribution \eqref{eq:noway} is certainly a challenge 
in many settings if the latent variable decomposing the sum is not to be used.

Since, in both \cite{scott:2002} and \cite{congdon:2006}, the $(\theta^{(t)})$'s are not simulated against
the correct target, the resulting averages of $P(M=k|y,\theta^{(t)})$, $\tilde{\varrho}_k(y)$
and $\hat{\varrho}_k(y)$, will both be biased, as demonstrated in the examples of Section \ref{sec:ex}.

\subsection{Improperty of the posterior}\label{sec:impro}
When resorting to the construction of pseudo-posteriors adopted by \cite{carlin:chib:1995}, 
\cite{congdon:2006} uses a {\em flat prior} as pseudo-prior on the parameters that are not in model $\mathfrak{M}_k$. 
More precisely, the joint prior distribution on $(\theta,M)$ is given by Congdon's (2006) formula (2),
\begin{align*}
P(\theta,M=k) &= \pi_k(\theta_k)\,\varrho_k\,\prod_{j\ne k} \pi(\theta_j|M=k) \\
 &= \pi_k(\theta_k)\,\varrho_k\,,
\end{align*}
which is indeed equivalent to assuming a flat prior as pseudo-prior on the parameters $\theta_j$
that are not in model $\mathfrak{M}_k$.

Unfortunately, this simplifying assumption has a dramatic consequence in that the 
corresponding joint posterior distribution of $\theta$ is never defined (as 
a probability distribution) since
$$
\pi(\theta|y) = \sum_{k=1}^D \pi_k(\theta_k|y)\,P(M=k|y)
$$
does not integrate to a finite value in any of the $\theta_k$'s (unless their support is compact).
While \cite{congdon:2006} states {\em that it is not essential that the priors for 
$P(\theta_{j\ne k}|M=k)$ are improper (p.348),} the truth is that they {\em cannot} be improper.

The fact that the posterior distribution on the saturated vector $\theta=(\theta_1,\ldots,\theta_D)$
does not exist obviously has negative consequences on the subsequent derivations, since a positive
recurrent Markov chain with stationary distribution $\pi(\theta|y)$ cannot be constructed. Similarly,
the fact that 
$$
P(M=k|y) = \int P(\theta,M=k|Y) \dd \theta
$$
does not hold any longer.

Note that \cite{scott:2002} does not follow the same track: when defining the pseudo-priors in his
formula (20), he uses the product definition\footnote{The indices on the priors have been added to make
notations 
consistent with the present paper.}
$$
P(\theta,M=k) = \pi_k(\theta_k)\,\varrho_k\,\prod_{\scriptstyle j\ne k} \pi_j(\theta_j)\,, \\
$$
which 
means that the true priors could also be used as pseudo-priors across all models.
However, we stress that \cite{scott:2002} does not refer to the construction of \cite{carlin:chib:1995}
in his proposal, nor does he use pseudo-priors in his simulations.

\subsection{Illustrations}\label{sec:ex}
We now proceed through several toy examples where all posterior quantities can 
be computed in order to evaluate the bias induced by both approximations and we observe
that, despite its theoretical bias, Congdon's (2006) can sometimes achieve a close
approximation of the posterior probability, but also that, in other settings, it 
may produce an unreliable evaluation.

\begin{figure}[hbt]
\centerline{\includegraphics[height=8truecm]{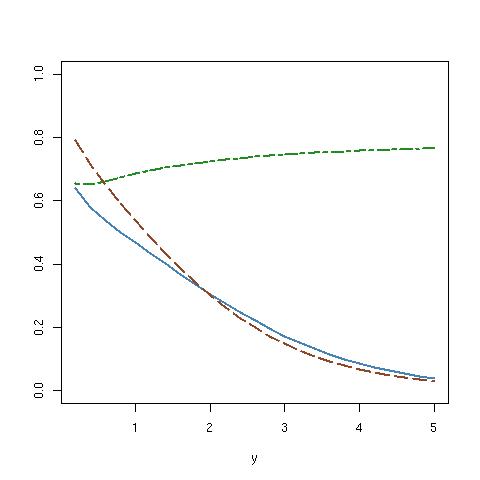}}
\caption{\label{fig:p1} {\bf Example \ref{ex:Uni}:}
Comparison of three approximations of $P(M=1|y)$ with the true value (in blue and full lines): 
Scott's (2002) approximation (in green and mixed dashes), Congdon's (2006) approximation (in brown and dashes),
while the correction of Scott's (2002) approximation is indistinguishable from the true value
(based on $N=10^6$ simulations).}
\end{figure}

\begin{ex}\label{ex:Uni}
Consider the case when a model $\mathfrak{M}_1:\,
y|\theta\sim\mathcal{U}(0,\theta)$ with a prior $\theta\sim\mathcal{E}xp(1)$ is
opposed to a model $\mathfrak{M}_2:\,y|\theta\sim\mathcal{E}xp(\theta)$ with a prior
$\theta\sim\mathcal{E}xp(1)$. We also assume equal prior weights on both models:
$\varrho_1=\varrho_2=0.5$.

The marginals are then
$$
m_1(y) = \int_y^\infty \theta^{-1}e^{-\theta}\dd \theta = \text{E}_1 (y)\,,
$$
where $\text{E}_1$ denotes the exponential integral function tabulated both in 
{\sf Mathematica} and in the {\sf GSL library}, and
$$
m_2(y) = \int_0^\infty \theta e^{-\theta (y+1)} \dd \theta = \frac{1}{(1+y)^2}\,.
$$
For instance, when $y=0.2$, the posterior probability of $\mathfrak{M}_1$ is thus equal to
\begin{align*}
P(M=1|y) &= m_1(y) / \{ m_1(y) + m_2(y) \} \\
         &= \text{E}_1 (y)/\{ \text{E}_1 (y) + (1+y)^{-2}\}\\ 
         &\approx 0.6378\,,
\end{align*}
while, for $y=0.9$, it is approximately $0.4843$. This means that, in the former case, the Bayes factor of $\mathfrak{M}_1$
against $\mathfrak{M}_2$ is $B_{12}\approx 1.760$, while for the latter, it decreases to $B_{12}\approx 0.939$.

The posterior on $\theta$ in model $\mathfrak{M}_2$ is a gamma $\mathcal{G}a(2,1+y)$ distribution and it can thus be simulated
directly. For model $\mathfrak{M}_1$, the posterior is proportional to $\theta^{-1}\,\exp(-\theta)$ for $\theta$
larger than $y$ and it can be simulated using a standard accept-reject algorithm based on an exponential $\mathcal{E}xp(1)$
proposal translated by $y$. 



Using simulations from the true (marginal) posteriors and the approximation of \cite{congdon:2006}, 
the numerical value of $\hat{\varrho}_1(y)$ based on $10^6$ simulations is
$0.7919$ when $y=0.2$ and $0.5633$ when $y=0.9$, which translates into Bayes factors of 
$3.805$ and of $1.288$, respectively. For the approximation of 
\cite{scott:2002}, the numerical value of $\tilde{\varrho}_1(y)$ is $0.6554$ (corresponding to a Bayes
factor of $1.898$) when $y=0.2$ and $0.6789$ when $y=0.9$ (corresponding to a Bayes factor of $2.11$), 
based on the same simulations.  Note that in the case $y=0.9$, a selection based on either 
approximation of the Bayes factor would select the wrong model.

If we use instead a correct simulation from the joint posterior \eqref{eq:noway}, which can be achieved by
using a Gibbs scheme with target distribution $P(\theta,M=k|y)$, 
we then get a proper MCMC approximation to the posterior probabilities
by the $\hat{P}(M=k|y)$'s. For instance, based on $10^6$ simulations, 
the numerical value of $\hat{P}(M=1|y)$ when $y=0.2$
is $0.6370$, while, for $y=0.9$, it is $0.4843$. Note that, due to the impropriety difficulty exposed in Section
\ref{sec:impro}, the equivalent correction for Congdon's (\citeyear{congdon:2006}) scheme cannot be implemented.

In Figure \ref{fig:p1}, the three approximations are compared to the exact value of $P(M=1|y)$ for a range of
values of $y$. The correct simulation produces a graph that is indistinguishable from the true probability, while
Congdon's (2006) approximation stays within a reasonable range of the true value and Scott's (2002) approximation
drifts apart for most values of $y$.
\end{ex}

The above correspondence of what is essentially Carlin and Chib's (1995) scheme
with the true numerical value of the posterior probability is obviously 
unsurprising in this toy example but more advanced setups see the approximation
degenerate, since the simulations from the prior are most often inefficient, especially
when the number of models under comparison is large. This is the reason why \cite{carlin:chib:1995}
introduced pseudo-priors that were closer approximations to the true posteriors.

The proximity of Congdon's (2006) approximation with the true value in Figure \ref{fig:p1} shows that
the method could possibly be used as a cheap first-order substitute of the true
posterior probability if the bias was better assessed. First, we note
that when all the componentwise posteriors are close to Dirac point masses at values $\hat\theta_k$,
Congdon's (2006) approximation is close to the true value
$$
\hat\varrho_k(y) \approx \varrho_k f_k(y|\hat\theta_k) \pi_k(\hat \theta_k)
\bigg/ \sum_{j=1}^D \left\{ \varrho_j\, f_j(y|\hat\theta_j) \pi_j(\hat\theta_j) \right\} \,.
$$
Further, the posterior expectation of $f_k(y|\theta_k^{(t)}) \pi_k(\theta^{(t)}_k)$
involves the integral of 
$$
\int \frac{f_k(y|\theta_k)^2 \pi_k(\theta_k)^2}{m_k(y)}\,\dd y\,,
$$ 
thus the bias is likely to be small in settings where the product $f_k(y|\theta_k^{(t)}) 
\pi_k(\theta^{(t)}_k)$ is peaked as in large samples, for instance. That the bias can almost 
completely disappear is exposed through a second toy example.

\begin{ex}\label{ex:No}
Consider the case when a normal model $\mathfrak{M}_1:\, y\sim\mathcal{N}(\theta,1)$ with a prior $\theta\sim\mathcal{N}(0,1)$
is opposed to a normal model $\mathfrak{M}_2:\,y\sim\mathcal{N}(\theta,1)$ with a prior $\theta\sim\mathcal{N}(5,1)$. We again
assume equal prior weights.

In that case, the marginals are available in closed form
$$
m_1(y) = \frac{1}{\sqrt{4\pi}}\,\exp-\frac{y^2}{4}
\quad\text{and}\quad
m_2(y) = \frac{1}{\sqrt{4\pi}}\,\exp-\frac{(y-5)^2}{4}
$$
and the posterior probability of model $\mathfrak{M}_1$ is
$$
P(M=1|y) = \left\{ 1 + \exp \frac{5(2y - 5)}{4}\right\}^{-1}\,.
$$
For argumentation's sake, assume that we now produce both sequences $(\theta_1^{(t)})$ and $(\theta_2^{(t)})$ 
from the posterior distributions $\mathcal{N}(y/2,1/2)$
and $\mathcal{N}((y+5)/2,1/2)$, respectively, by using the {\em same sequence} of $\epsilon_t\sim\mathcal{N}(0,1)$, i.e.
$$
\theta_1^{(t)}=\frac{y}{2}+\frac{1}{\sqrt{2}}\epsilon_t
\quad\text{and}\quad
\theta_2^{(t)}=\frac{y+5}{2}+\frac{1}{\sqrt{2}}\epsilon_t\,.
$$
Using those sequences, we then obtain that
\begin{align*}
\frac{
\exp -\frac{1}{2} (y-\theta_1^{(t)})^2-\frac{1}{2}(\theta_1^{(t)})^2
}{
\exp -\frac{1}{2} (y-\theta_2^{(t)})^2-\frac{1}{2}(\theta_2^{(t)}-5)^2
} &=
\frac{
\exp -\frac{1}{2} (y-\frac{y}{2}-\frac{1}{\sqrt{2}}\epsilon_t)^2-\frac{1}{2}(\frac{y}{2}+\frac{1}{\sqrt{2}}\epsilon_t)^2
}{
\exp -\frac{1}{2} (y-\frac{y+5}{2}-\frac{1}{\sqrt{2}}\epsilon_t)^2-\frac{1}{2}(\frac{y+5}{2}+\frac{1}{\sqrt{2}}\epsilon_t-5)^2
}\\
&= 
\frac{
\exp -\frac{1}{2} (\frac{y}{2}-\frac{1}{\sqrt{2}}\epsilon_t)^2-\frac{1}{2}(\frac{y}{2}+\frac{1}{\sqrt{2}}\epsilon_t)^2
}{
\exp -\frac{1}{2} (\frac{y-5}{2}-\frac{1}{\sqrt{2}}\epsilon_t)^2-\frac{1}{2}(\frac{y-5}{2}+\frac{1}{\sqrt{2}}\epsilon_t)^2
}\\
&= \exp -\frac{5}{4}(2y-5)\,,
\end{align*}
independently of $\epsilon_t$, and thus that Congdon's (2006) approximation is truly exact using this device!
Figure \ref{fig:No} shows the difference due to using two independent sequences of $10^4$ $\epsilon_t$'s [instead
of one single sequence] and the severe discrepancy resulting from Scott's approximation. (Note that using an artificial
MCMC sampler in this case would only increase the variability of the approximations.)
\begin{figure}[hbtp]
\centerline{\includegraphics[height=8truecm]{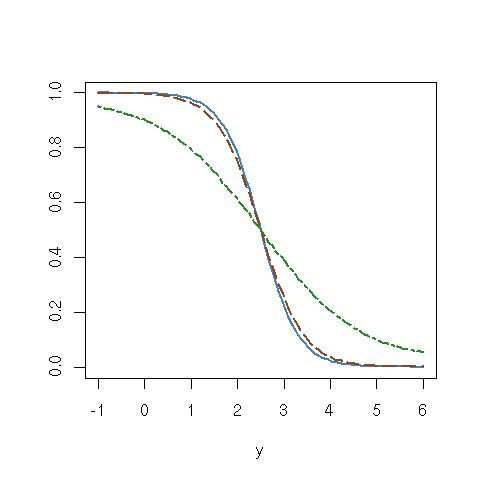}}
\caption{\label{fig:No} {\bf Example \ref{ex:No}:}
Comparison of two approximations of $P(M=1|y)$ with the true value (in blue and full lines):
Scott's (2002) approximation (in green and mixed dashes) and Congdon's (2006) approximation (in brown and long dashes)
(based on $N=10^4$ simulations).}
\end{figure}
\end{ex}

The approximation may also be rather crude, as shown in the following example,
inspired from an example posted on Peter Congdon's web-page in connection with \cite{congdon:2006b}.

\begin{ex}\label{ex:Bin}
Consider comparing $\mathfrak{M}_1:\,y\sim\mathcal{B}(n,p)$ when $p\sim\mathcal{B}e(1,1)$ with 
$\mathfrak{M}_2:\,y\sim\mathcal{B}(n,p)$ when $p\sim\mathcal{B}e(m,m)$. Once again, the posterior
probability can be computed in closed form since the Bayes factor is given by
$$
B_{12}  = \frac{(n+1)!}{y!(n-y)!}\,\frac{(m+y-1)!(m+n-y-1)!}{(m+n-1)!}\,\frac{(m-1)!^2}{(2m-1)!}\,.
$$
The simulations of $p_1^{(t)}$ from the posterior $\mathcal{B}e(y+1,n-y+1)$ in model $\mathfrak{M}_1$ and of
$p_2^{(t)}$ from the posterior $\mathcal{B}e(y+m,m+n-y)$ in model $\mathfrak{M}_2$ are straightforward (and
obviously do not require an extra MCMC step). Figure \ref{fig:Bin} 
shows the impact of Congdon's (2006) approximation on the evaluation
of the posterior probability for $n=15$ and $m=100$: the magnitude is the same but, in that
case, the numerical values are quite different.
\begin{figure}[hbtp]
\centerline{\includegraphics[height=8truecm]{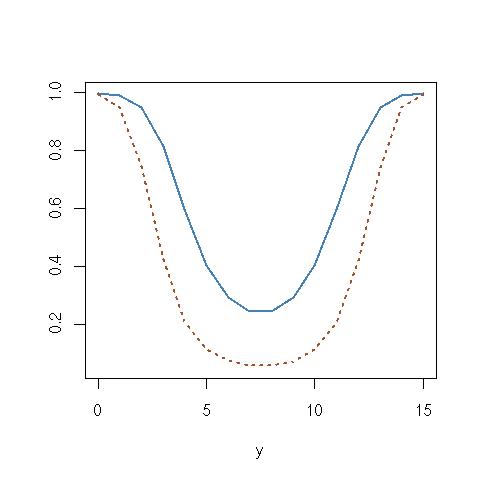}}
\caption{\label{fig:Bin} {\bf Example \ref{ex:Bin}:}
Comparison of Congdon's (2006) (in brown and dashed lines) approximation of $P(M=1|y)$ with the true value 
(in blue and full lines) when $n=15$ and $m=510$ (based on $N=10^4$ simulations).}
\end{figure}

In the case of three models in competition, namely when $y\sim\mathcal{B}(n,p)$
and the three priors are $p\sim\mathcal{B}e(1,1)$,
$p\sim\mathcal{B}e(a,b)$ and $p\sim\mathcal{B}e(c,d)$, the differences may be of the same order, as shown in
Figure \ref{fig:Bin3}, but the discrepancy is nonetheless decreasing with the sample size $n$.
\begin{figure}[hbtp]
\centerline{\includegraphics[height=12truecm]{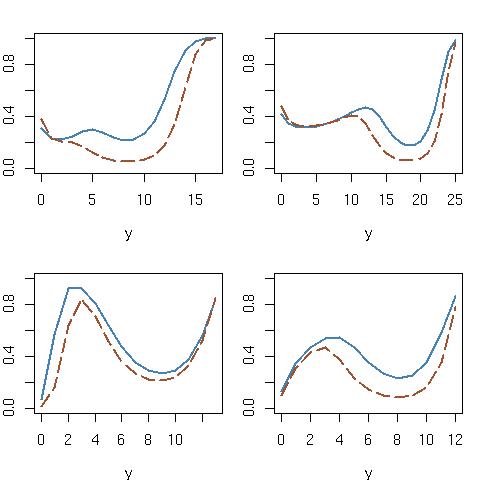}}

\caption{\label{fig:Bin3} {\bf Example \ref{ex:Bin}:}
Comparison of Congdon's (2006) (in brown and dashed lines) approximation of $P(M=1|y)$ with the true value
(in blue and full lines) when $(n,a,b,c,d)$ is equal to $(17,2.5,12.5,501.5,500)$,
$(25,1.5,4,540,200)$, $(13,.5,100.5,20,10)$ and $(12,.3,1.8,200,200)$, respectively
(based on $N=10^4$ simulations).}
\end{figure}
\end{ex}

At last, the approximation may fall very far from the mark, as demonstrated in the following example
where the approximation has an asymptotic behaviour opposite to the one of the true posterior probability.
\begin{ex}\label{ex:Ex}
Consider comparing $\mathfrak{M}_1:\,y\sim\mathcal{N}(0,1/\omega)$ with $\omega\sim\mathcal{E}xp(a)$
against $\mathfrak{M}_2:\,\exp(y)\sim\mathcal{E}xp(\lambda)$ with $\lambda\sim\mathcal{E}xp(b)$.
The corresponding marginals are given in closed form by
$$
m_1(y)=\int_0^\infty \sqrt{\frac{\omega}{2\pi}}\,e^{-(y^2/2)\omega}\,ae^{-a\omega}\,\text{d}\omega 
= \frac{a}{\sqrt{2\pi}}\,\frac{\Gamma(3/2)}{(a+y^2/2)^{3/2}}
$$
and
$$
m_2(y)= \int_0^\infty e^y\,\lambda e^{-e^y\lambda}\,be^{-b\lambda}\,\text{d}\lambda
= \frac{b\,e^y}{(b+e^y)^2}\,.
$$
The associated posteriors are $\omega|y\sim\mathcal{G}a(3/2,a+y^2/2)$ and $\lambda|y\sim\mathcal{G}a(2,b+e^y)$.
Figure \ref{fig:Ex} shows the comparison of the true posterior probability of $\mathfrak{M}_1$
with the approximation for various values of $(a,b)$
and it indicates a very poor fit when $y$ goes to $+\infty$. 

It is actually possible to show that the approximation
always converges to $0$ when $y$ goes to $+\infty$, while the true posterior probability goes to $1$. Indeed,
when $y$ goes to $+\infty$, the Bayes factor is
$$
\frac{m_1(y)}{m_2(y)} \approx \frac{a\Gamma(3/2)}{b\sqrt{2\pi}}\,\frac{e^{2y}}{e^y(y^2/2)^{3/2}}\,,
$$
which goes to $+\infty$ while, since $\omega^{(t)}=\epsilon_t/(a+y^2/2)$ and $\lambda^{(t)}=\upsilon_t/(b+e^y)$,
with $\epsilon_t\sim\mathcal{G}(3/2,1)$ and $\upsilon_t\sim\mathcal{G}(2,1)$,
$$
\frac{f_1(y|\omega^{(t)})\pi_1(\omega^{(t)})}{f_2(y|\lambda^{(t)})\pi_2(\lambda^{(t)})}=
\frac{a}{b\sqrt{2\pi}}\,\frac{\sqrt{\epsilon_t}e^{-\epsilon_t}}{\upsilon_te^{-\upsilon_t}}\,
\frac{b+e^y}{e^y(a+y^2/2)^{1/2}}\approx
\frac{a}{b\sqrt{2\pi}}\,\frac{\sqrt{\epsilon_t}e^{-\epsilon_t}}{\upsilon_te^{-\upsilon_t}}\, \frac{\sqrt{2}}{y}\,,
$$
which goes to $0$ for all $(\epsilon_t,\upsilon_t)$. The discrepancy is then extreme.
\begin{figure}[hbtp]
\centerline{\includegraphics[height=12truecm]{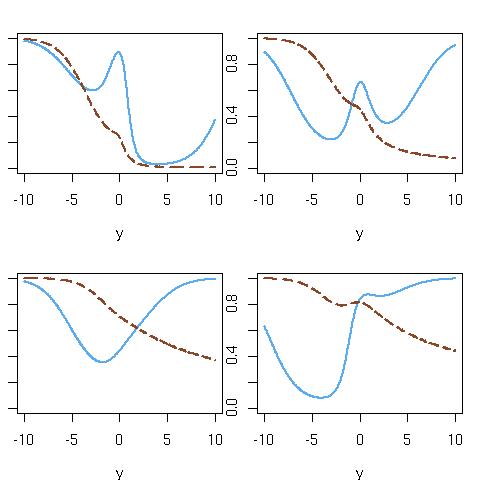}}

\caption{\label{fig:Ex} {\bf Example \ref{ex:Ex}:}
Comparison of Congdon's (2006) (in brown and dashed lines) approximation of $P(M=1|y)$ with the true value
(in blue and full lines) when $(a,b)$ is equal to $(.24,8.9)$,
$(.56,.7)$, $(4.1,.46)$ and $(.98,.081)$, respectively
(based on $N=10^4$ simulations).}
\end{figure}
\end{ex}

\section*{Acknowledgements}

Both authors are grateful to Brad Carlin and to the editorial board for helpful suggestions and to Antonietta 
Mira for providing a perfect setting for this work during the ISBA-IMS ``MCMC'ski 2" conference in Bormio,
Italy. The second author is also grateful to Kerrie Mengersen for her invitation to ``Spring Bayes 2007"
in Coolangatta, Australia, that started our reassessment of those papers. This work had been 
supported by the Agence Nationale de la Recherche (ANR, 212, rue de Bercy 75012 Paris)
through the 2005-2008 project {\sf Adap'MC}.

\bibliographystyle{ba}
\bibliography{biblio.bib}

\end{document}